\documentclass[amsmath,amssymb,aps,prl,reprint,floatfix,superscriptaddress]{revtex4-2}

\usepackage{listings}
\usepackage{graphicx}
\usepackage{dcolumn}
\usepackage{xcolor}
\usepackage{bm}
\usepackage[version=4]{mhchem}
\usepackage[breaklinks,colorlinks,linkcolor={blue}, %
            citecolor={blue},urlcolor={blue}]{hyperref}
\usepackage[normalem]{ulem} 
\begin{document}

\title{$\Delta$SCF in \texttt{VASP} for excited-state defect computations: tips and pitfalls}
\author{Yihuang Xiong} 
\affiliation{Thayer School of Engineering, Dartmouth College, Hanover, New Hampshire 03755, USA}
\author{Geoffroy Hautier} 
\affiliation{Thayer School of Engineering, Dartmouth College, Hanover, New Hampshire 03755, USA}
\date{\today}

\begin{abstract}
$\Delta$SCF with constrained occupations have been wildly used to investigate the excited-state and optical properties of defects. Recent studies have demonstrated that combining $\Delta$SCF with hybrid functionals yields good accuracy in predicting defect properties. The Vienna Ab initio Simulation Package (\texttt{VASP}) is one of the most widely used quantum mechanical packages based on plane-wave methods. Despite the increasing application of $\Delta$SCF as implemented in \texttt{VASP} for defect studies, detailed walkthroughs explaining how to conduct these calculations remain limited, making this approach a nontrivial task. Applying $\Delta$SCF with hybrid functionals can present convergence challenges; worse, it may sometimes converge to incorrect excited states and can go largely unnoticed. This document aims to serve as a concise guide outlining what we think might be the appropriate approach for performing $\Delta$SCF calculations in \texttt{VASP}. We benchmark this method by simulating excited states for a particularly challenging system: the neutral charge state of the silicon vacancy (SiV$^0$) defect in diamond. By highlighting potential pitfalls, we hope this document encourages further discussion within the community and assists researchers experiencing difficulties with this technique. The guidelines provided here are largely based on private discussions with Oscar Bulancea Lindvall from Link{\"o}ping University and Chris Ciccarino from Stanford University.
\end{abstract}

\maketitle

\section{Introduction}
Defects are ubiquitous in semiconductors and play a pivotal role in thermodynamic, electronic, and optical properties. The resurgence of employing defects as solid-state qubits has attracted great attention for quantum applications including spin-photon interfaces. One critical requirement for spin-photon interfaces is that the quits can be initialized and readout using light. Density functional theory has been successful in predicting ground-state properties, and by constraining occupations (the so-called $\Delta$SCF method), its applications have been extended to excited-state properties such as vertical excitation energy (VEE), zero-phonon line (ZPL), and photoluminescence lineshape, etc..\cite{Alkauskas_2014,Alkauskas2016}, making it a powerful method for studying defects. Moreover, when combined with hybrid functionals such as HSE06, $\Delta$SCF has demonstrated strong predictive capabilities.

Unfortunately, performing $\Delta$SCF can be be a nontrivial task, often plagued by convergence issues or unintended convergence to incorrect excited states. In this document, we will illustrate the procedure for applying the $\Delta$SCF method combined with hybrid functionals using the neutral silicon split-vacancy defect in diamond (SiV$^0$) as an example. Specifically, we aim to address three primary challenges encountered in $\Delta$SCF calculations:
\begin{itemize}
  \item Basic $\Delta$ SCF INCAR settings.
  \item \texttt{\texttt{VASP}} version.
  \item Restarting strategies.
\end{itemize}
All the inputs and outputs of the following calculations are available at
\url{https://github.com/defectmrE/DSCF_VASP}.
\section{Results}
\begin{figure}[t]
 	\centering
 	\includegraphics[width=0.45\textwidth]{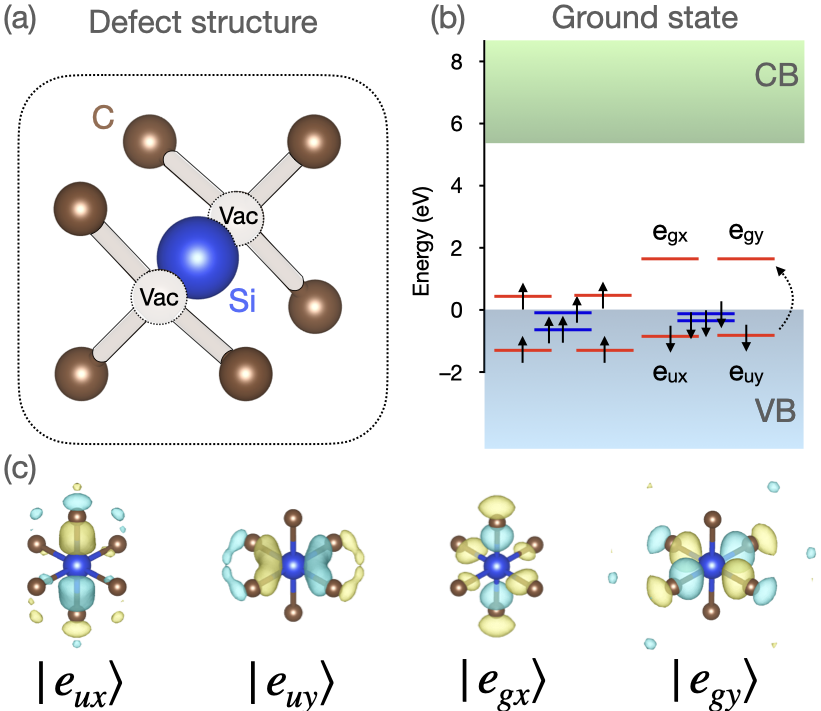}    
 	\caption{(a) SiV$^0$ defect structure in $D_{3d}$ symmetry and (b) Kohn-Sham defect level of the ground state computed using HSE06. Defect level and valence band maximum are labeled with red and blue lines, respectively. The noticeable amount of the splits of valence bands is due to the finite size effect that caused strain in the cell. The order of defect orbital labeling is illustrative, the specific order depends on the very calculations. (c) Wavefuncitons of the defect orbitals. 
 	}
 	\label{siv.struct}
\end{figure}

SiV$^0$ in diamond exhibits a defect point group of $D_{3d}$, as illustrated in Figure.\ref{siv.struct}(a). The defect level diagram computed at HSE level is shown in Figure.\ref{siv.struct}(b). It is important to note that the $e_{ux}$ and $e_{uy}$ levels (Figure.\ref{siv.struct}(c)) are resonant within the VBM, which makes simulating its excited states properties challenging. We will discuss this in detail below.

\begin{figure*}[t]
 	\centering
 	\includegraphics[width=0.65\textwidth]{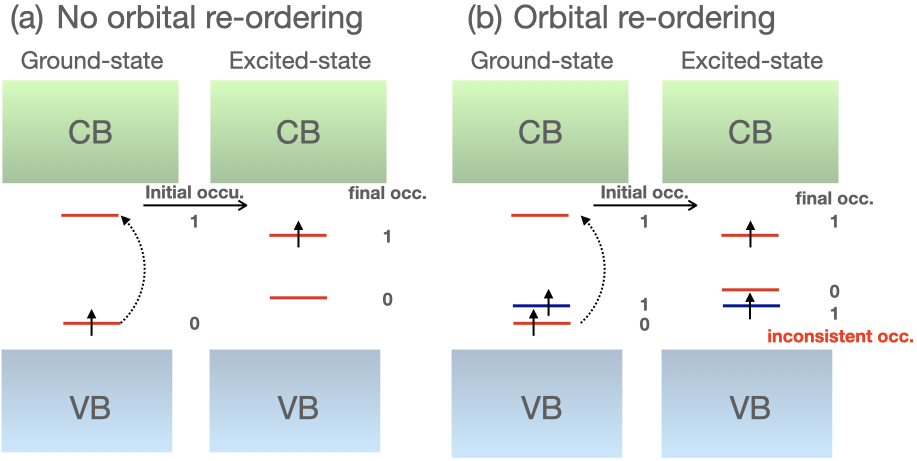}    
 	\caption{(a) $\Delta$SCF exited states where the relaxation does not result in orbital reordering, the occupation are consistent before and after. (b) $\Delta$SCF yields reordering of orbitals and results in inconsistency in occupations, prone to electronic step divergence or converge to undesired excited states.
 	}
 	\label{orbital_ordering}
\end{figure*}

Many factors can influence the convergence $\Delta$SCF calculations, among which orbital re-ordering can be particularly challenging to address. Figure.\ref{orbital_ordering} shows the case where orbital reordering does \textbf{not} occur: upon promoting one electron to a higher-energy orbital, the occupied orbital energy decreases, while the unoccupied orbital energy slightly increases. Crucially, no crossing occurs between these two orbitals, and thus the initial orbital occupations resemble the final ones. Such a scenario is generally considered ``well-behaved'', allowing calculations to proceed from scratch without restarting from a pre-generated wavefunction or applying orbital-ordering constraints (tag $\textbf{LDIAG}$)\cite{Dhaliah_PRM_2022,Xiong_jacs_2024}.

However, for situation b in Figure.\ref{orbital_ordering}, things are getting tricky. Due the populating/depopulating orbitals, the occupancy has changed \cite{Xiong_midgap, Li2023Solid, thiering2019}, and this is likely to cause convergence issues or converged to undesired states. It can be seen that SiV$^0$ belongs to situation 2, where promoting an electron from $e_u$ to $e_g$ involves crossing over three VBM states of diamond. 

\section{The scope of this document: What calculations are discussed?}

We aim to apply $\Delta$SCF with HSE06. This choice is motivated not only by the improved accuracy of hybrid functionals in predicting energies and electronic structures, but also because semilocal functionals such as PBE are less prone to orbital reordering issues. Typically, the shifts in defect orbital energies are much milder in PBE calculations, potentially making them less susceptible to scenario (b). Nonetheless, it remains essential to verify that the final results indeed correspond to the intended excited states.

\section{Basic INCAR setup}
\textbf{KEY TAKEAWAY: LDIAG=False}.
Here is an example of a few important tags in the calculations:
\begin{lstlisting}[language={},basicstyle=\ttfamily,frame=single]
ALGO = ALL
LDIAG = FALSE
ISMEAR = -2
FERDO = 425*1 1*0 4*1 178*0
FERWE = 431*1 177*0
\end{lstlisting}
The most important tag here is $\textbf{LDIAG}$, which is set to $\textbf{True}$ by default. By explicitly setting it to $\textbf{True}$ one can prevent orbital reordering, thus achieving the desired electronic structures.  Note that this tag only works with $\textbf{ALGO=All/Damp}$ for hybrid functional calculations, and $\textbf{ALGO=All}$ is selected here for faster electronic convergence. We also want to point out that Lindvall and the coauthors recently reported that the combination of $\textbf{ALGO=Damp}$ and $\textbf{LDIAG=False}$ in \texttt{VASP.5.4.1} was the only combination that yielded correct $\Delta$SCF results.~\cite{2024_BulanceaLindvall}, further details can be found in their Appendix. 

\section{\texttt{VASP} version}
\textbf{KEY TAKEAWAY: \texttt{VASP.5.4.4}}.
Ideally, software version differences would not affect functionality; unfortunately, this is not the case here. In our benchmark study, we consider the following \texttt{VASP} versions:

\begin{itemize}
    \item \texttt{VASP.5.4.4}
    \item \texttt{VASP.5.4.4}-patched
    \item \texttt{VASP.6.2.1}
    \item \texttt{VASP.6.4.2/6.4.3}
\end{itemize}

The patch proposed by Lindvall and coauthors \cite{2024_BulanceaLindvall} targets main.F and can be found at \url{https://github.com/defectmrE/DSCF_VASP} or obtained by contacting Lindvall. Through testing, we identified a known issue in \texttt{VASP.6.2.1} with the tag $\textbf{ISMEAR=-2}$, in which occupation constraints are inadvertently removed during ionic relaxation, causing the electronic structure to revert to its ground state. Additionally, \texttt{VASP.6.4.2}  and \texttt{6.4.3} experience convergence difficulties despite exhaustive tests using various $\textbf{TIME}$ parameters and charge-mixing schemes. We suspect that the $\textbf{LDIAG=False}$ setting is ineffective in these \texttt{VASP} 6.x versions. Consequently, this document focuses exclusively on \texttt{VASP.5.4.4}  and its patched variant.

\section{Restart vs. from scratch}
\textbf{KEY TAKEAWAY: \texttt{VASP}.5.4.4-patched, restart from PBE wavefunction}.

\begin{figure}[t]
 	\centering
 	\includegraphics[width=0.45\textwidth]{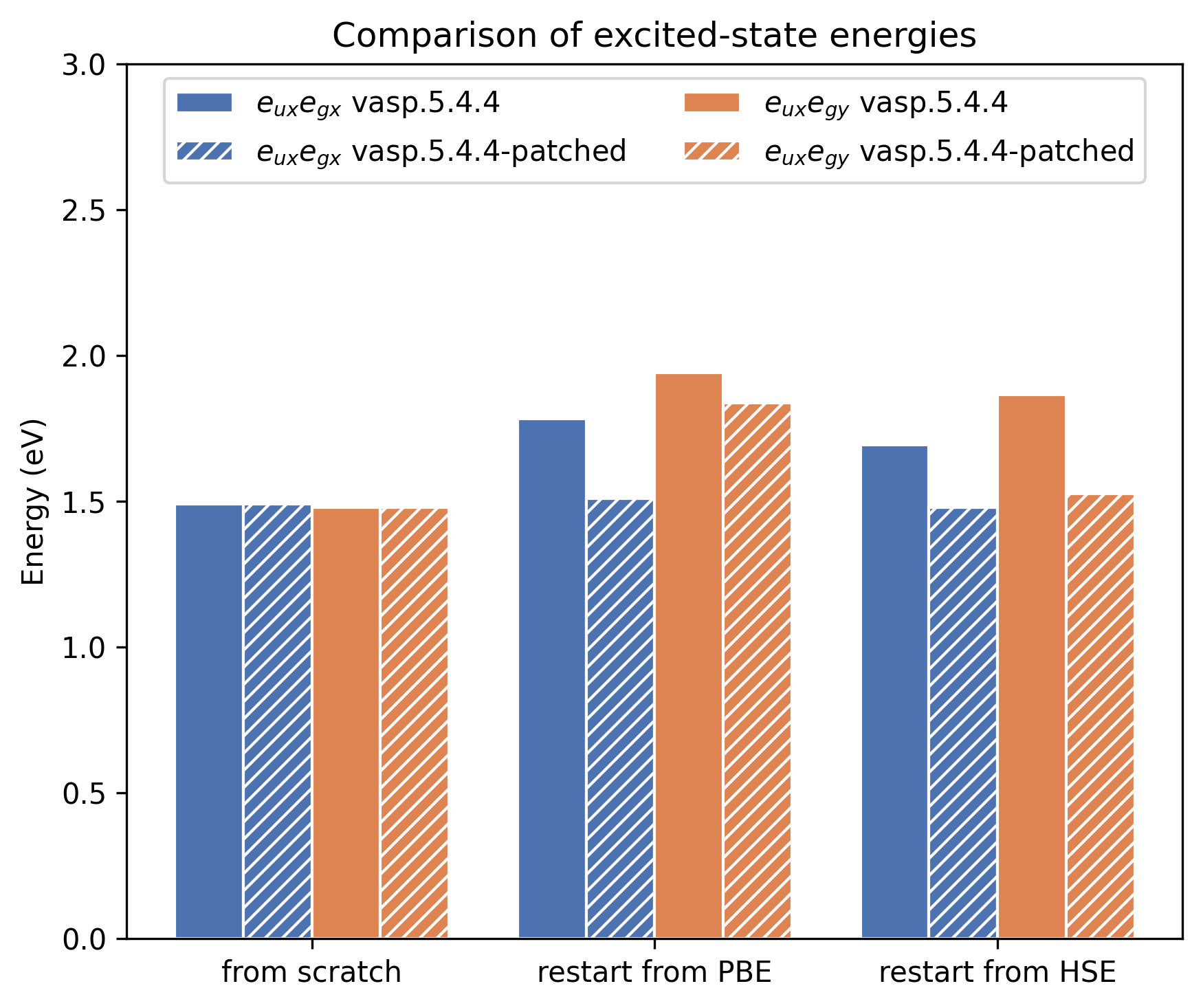}    
 	\caption{Comparison of ZPLs computed using $\Delta$SCF methods implemented in \texttt{VASP.5.4.4} and \texttt{VASP.5.4.4}-patched. The two-hole wavefunctions are constrained on $e_{ux}^0e_{gx}^0$ and $e_{ux}^0e_{gy}^0$ ($\mathcal{A} \lvert e_{ux} e_{gx} \rangle$ and $\mathcal{A} \lvert e_{ux} e_{gy} \rangle$), respectively. 
 	}
 	\label{zpls}
\end{figure}

We tested three strategies for performing $\Delta$SCF calculations from the HSE-optimized ground-state structures: \textbf{1)} from scratch, \textbf{2)} restart from PBE wavefunction, and \textbf{3)} from the HSE wavefunction. Our goal is to simulate excitations corresponding to $e_{ux}^1e_{uy}^0e_{gx}^1e_{gy}^0$ and $e_{ux}^1e_{uy}^0e_{gx}^0e_{gy}^1$ in the spin-minority channel (denoted as $\mathcal{A} \lvert e_{uy} e_{gy} \rangle$ and $\mathcal{A} \lvert e_{uy} e_{gx} \rangle$ for two-hole wavefunction, follwoing the naming convention of Ref.\cite{thiering2019}). As previously discussed, SiV$^0$ is a challenging system due to \textbf{1)} the emptied $e_{gx}/e_{gy}$ states will pop above the valence bands and results in orbital reordering, and \textbf{2)} the degeneracy between $e_{ux}$/$e_{uy}$ and $e_{gx}$/$e_{gy}$ requires precise constraints on the desired orbitals. The ZPLs simulations are presented in Figure~\ref{zpls}. 

Before presenting the results, we emphasize that performing calculations from scratch can be challenging. Setting initial occupations based solely on ground-state electronic structures does not guarantee convergence to the correct excited-state configuration because of orbital reordering issues. Consequently,  multiple calculations with different initial occupations might be required to identify those that yield the desired excited states. Unfortunately, there are situations where the targeted excited states can not be achieved though many different occupations are tested.

The ZPL calculations are performed without symmetry constraints to include Jahn-Teller (JT) effects, with the goal of differentiating JT energies from distinct orbitals. The results are summarized in Figure.\ref{zpls}. Previously, Thiering and Ciccarino reported $E_{\mathrm{JT}}$ for the $\mathcal{A} \lvert e_{ux} e_{gx} \rangle$ and $\mathcal{A} \lvert e_{ux} e_{gy} \rangle$ of 258 meV in a 512-atom supercell\cite{thiering2019}. For calculations starting \textbf{from scratch}, the ZPLs fail to reflect the difference of $E_{\mathrm{JT}}$ with underestimated values, irrespective of whether \texttt{VASP.5.4.4} is patched or not. For calculations restarting from PBE wavefunctions, \texttt{VASP.5.4.4} predicts higher ZPLs compared to the patched version. On the contrary, the \texttt{VASP}-5.4.4-patched version shows the right order that $e_{ux}^0e_{gx}^0$ has a stronger JT relaxation compared to that of $e_{ux}^0e_{gy}^0$. We observed the energy difference to be 328 meV in this case. This value is larger than the reported values of 258 meV due to the smaller cell size that is used in this study.

Moving on to the calculations restarted from $\textbf{HSE}$ wavefunction, again, \texttt{VASP.5.4.4} predicts a too high of ZPLs, and \texttt{VASP.5.4.4}-patched failed to recognize the energy differences of $E_{\mathrm{JT}}$ with a small $E_{\mathrm{JT}}$. Therefore, we conclude that \textbf{\texttt{VASP.5.4.4}-patched in combination with restarting from PBE wavefunctions, constrained occupations, $\textbf{ALGO=ALL}$ and $\textbf{LDIAG=FALSE}$} yields the right $\Delta$SCF excitations. We will discuss how to verify if the wavefunctions obtained using $\Delta$SCF are indeed the ground-state with respect to its the constrained occupations in the next section.

It is important to note that the abnormal ZPLs could originate from wavefunciton distortions introduced by $\textbf{LDIAG=FALSE}$, which have previously been reported by Sajid and coauthors. They found that such wavefunctions might represent unconverged mimics of lower-energy states, resulting in artificially high excitation energies. They proposed an iterative approach of toggling  $\textbf{LDIAG}$ and updating occupations through $\textbf{FERDO}$ and $\textbf{FERDO}$ until the desired excited states are achieved\cite{reimers2018,Reimers2020,ali2019colour}. Similar methodologies have been developed and employed by Ciccarino in his work\cite{Ciccarino_npj_2020}. Although these iterative methods can successfully yield the final excited states, preliminary tests indicate they require at least five times the computational resources compared to Lindvall’s patched method. Consequently, the patch proposed by Lindvall and coauthors currently represents the most efficient approach for accurately calculating excited-state energies in \texttt{VASP}.

We also note that other approaches exist to circumvent these issues, or overcome these problems using higher level theory. For instance, TDDFT with forces have been developed by Yu as implemented in WEST\cite{Jin.JCTC.2023}, CP2K\cite{Hehn2022}, and direct-optimization of maximally overlap orbitals (DO-MOM) as implemented in GPAW\cite{2021_Ivanov, 2020_Levi}. Benchmarking these alternative methods, however, is beyond the scope of this document.”

\section{Validation of the excited states}
We can verify the validity of the final energies and wavefunctions by lifting the order constraints imposed by the $\textbf{LDIAG=False}$. It can be performed as follows: first, taking the excited-state wavefunction (WAVECAR) and reorder it based on the eigenvalues (EIGENVAL) in ascending order; Next, update the $\textbf{FERWE}$ and $\textbf{FERDO}$ according to the updated wavefunctions; Finally, restart the calculation from the reordered wavefunction, if the calculation converges in a single step (both the electronic and ionic steps), it indicates that the ground-state wavefunction has been achieved and confirm the correct convergence. We have verified this for all the calculations and only the ones from \texttt{VASP.5.4.4}-patched-PBE-restart calculations fulfill such requirement. The WAVECAR ordering script by Chris Ciccarino is also available at \url{https://github.com/defectmrE/DSCF_VASP}.

\section{Conclusions}

In this document, we benchmarked the $\Delta$SCF method as implemented in \texttt{VASP}. After systematically testing various input settings, software versions, and restarting approaches, we conclude that the optimal procedure \textbf{\texttt{VASP.5.4.4}-patched version when restart from PBE wavefunctions, in combination of the constrained occupations and $\textbf{ALGO=ALL}$ and $\textbf{LDIAG=FALSE}$}(as proposed by Lindvall\cite{2024_BulanceaLindvall}). We hope that the issues related to the $\Delta$SCF behavior will be resolved by the \texttt{VASP} developers in future releases, alongside the GPU-accelerated hybrid functional implementations to enhance the efficiency and scalability of $\Delta$SCF calculations for quantum defect studies.

\section{Computational methods}
All the first-principles calculations were performed using \texttt{VASP}~\cite{G.Kresse-PRB96,G.Kresse-CMS96} and the projector-augmented wave (PAW) method~\cite{P.E.Blochl-PRB94}. Spin-polarized computations are performed in a supercell with 215 atoms and relaxed with a fixed volume until the ionic forces are smaller than 0.01~eV/\r{A}. The Brillouin zone is sampled with $\Gamma$ point. A cutoff energy of 400 eV is applied. All computations were performed using the Heyd-Scuseria-Ernzerhoff (HSE)~\cite{Heyd2003} functional with 25\% exact exchange. \texttt{VASP.5.4.4/6.2.1/6.4.2/6.4.3} are tested in this work.

\section{Acknowledgments}
\begin{acknowledgments}
We gratefully acknowledge Oscar Bulancea Lindvall and Christopher J. Ciccarino for generously sharing their insights and for engaging in fruitful discussions.
\end{acknowledgments}

\end{document}